\documentclass[twocolumn,prb,showpacs,preprintnumbers,amsmath,superscriptaddress,amssymb]{revtex4}

\usepackage{graphicx}
\usepackage{dcolumn}

\begin{document}

\title{Effective Operator for $dd$ Transitions in Nonresonant Inelastic X-ray Scattering}

\author{Michel van Veenendaal\cite{ESRF}}
  \affiliation{Dept. of Physics, Northern Illinois University, De Kalb, Illinois 60115}
  \affiliation{Advanced Photon Source, Argonne National Laboratory, 9700 S. Cass Avenue, Argonne, Illinois 60439}
\author{M. W. Haverkort}
  \affiliation{ II. Physikalisches Institut, Universit\"at zu K\"oln, Z\"ulpicher Str. 77, D-50937, K\"oln, Germany}
  \affiliation{Max Planck Institute for Solid State Research, Heisenbergstra{\ss}e 1, D-70569 Stuttgart Germany }

\date{\today}

\begin{abstract}
Recent experiments by Larson {\it et al.} \cite{Larson} demonstrate the 
feasibility of measuring local $dd$ excitations using nonresonant
inelastic X-ray scattering (IXS). We establish a general framework for the
interpretation where the $dd$ transitions created in the scattering process
 are expressed in effective one-particle operators that 
follow a simple selection rule.  The different operators can 
be selectively probed by employing 
their different dependence on the direction and magnitude 
of the transferred momentum.
We use the operators to explain the presence of nodal directions and the 
nonresonant IXS in specific directions and planes. We demonstrate 
how nonresonant IXS  can be used to extract valuable ground state information
for orbiton excitations in manganite.
\end{abstract}

\pacs{ 78.70.Ck, 71.20.Be} 

\maketitle

\section{ Introduction} Transition-metal compounds display a
wide variety of exciting phenomena such as high-$T_c$ superconductivity,
colossal magneto resistance, and metal-insulator transitions resulting
 from the strong interplay between the
charge, spin, and orbital degrees of freedom. 
Knowledge  of the electronic and magnetic structure 
can be obtained by a variety of spectroscopic techniques, such as
optical spectroscopy, photoemission, X-ray absorption. 
A general understanding of spectroscopy and its uses 
has always been a crucial aspect in
the advancement of condensed-matter physics.
Recently, inelastic X-ray scattering (IXS),  both on and off resonance,
has attracted considerable attention. 
Resonant IXS (RIXS) \cite{Kotani} is a second-order process
dominated by ${\bf p}\cdot {\bf A}$, where ${\bf p}$ is the momentum
and ${\bf A}$ the vector potential. The incoming X ray excites an electron
from a deep-lying core state into the valence shell, and one measures the
radiative decay of the core hole.
 For RIXS at the transition-metal $L$ and $M$ edges, 
the dipole transitions create $dd$ excitations
that include spin-flips due to the strong intermediate-state 
spin-orbit coupling \cite{Butorin,MvVRIXSLM}. The transitions
can be described with an effective operator approach 
\cite{MvVRIXSLM} in the fast-collision approximation. RIXS at the $K$ edge, 
where excitations are predominantly shake-up processes 
by the strong $1s$ core potential \cite{Hasan},  can be
related in certain limits to the dynamic structure factor $S_{\bf q}(\omega)$
\cite{vandenBrinkRIXS}. 

Nonresonant IXS, on the other hand, involves the interaction
\begin{eqnarray}
\frac{e^2}{2m} {\bf A}^2=\frac{e^2}{2m} {\bf e}'\cdot{\bf e}
e^{i{\bf q}\cdot {\bf r}},
\label{A2}
\end{eqnarray}
 with ${\bf q}={\bf k}\mathopen{-}{\bf k}'$
where ${\bf e}/{\bf e}'$ and ${\bf k}/{\bf k}'$
are the polarization and wave vectors of the incoming/outgoing X rays,
respectively. The interaction due to the $A^2$ term is in principle weak.
However, the successful experiments by Larson et al. \cite{Larson} 
demonstrate the feasibility of measuring local $dd$ excitations and 
comparison of the radial matrix elements indicate that these transitions
can in principle also be observed in other transition-metal 
and rare-earth compounds.
The appealing feature of nonresonant IXS is
that the cross section  is proportional to the dynamic structure
factor: $\frac{d^2\sigma}{d\Omega d\omega}\sim S_{\bf q}(\omega)$, which
through the fluctuation-dissipation theorem is connected to 
the imaginary part of the density response function,
\begin{eqnarray}
\chi_{\bf q}(\omega)=\langle g|\rho_{\bf q}
\frac{1}{\omega+E_g-H+i0^+} \rho_{-{\bf q}} |g\rangle,
\end{eqnarray}
with  $E_g$ the ground state energy and 
where the Hamiltonian $H$, for transition-metal 
compounds,  includes strong many-body interactions; $\rho_{\bf q}$ 
is the density  operator 
\begin{eqnarray}
\rho_{\bf q}=\sum_{{\bf k}nn' \sigma}
\langle \psi_{{\bf k}+{\bf q},n'\sigma} |e^{i{\bf q}\cdot{\bf r}}
|\psi_{{\bf k}n\sigma}\rangle
c^\dagger_{{\bf k}+{\bf q},n'\sigma} c_{{\bf k}n\sigma}
\end{eqnarray}
 where  $c_{{\bf k}n\sigma}^\dagger$ creates an  electron in a state
$\psi_{{\bf k}n\sigma}$ 
where $n$  is the band index and $\sigma=\uparrow,\downarrow$. 
Without matrix elements $\rho_{\bf q}$ becomes equivalent to the 
charge density operator $\rho_{\bf q}=\sum_{{\bf k}\sigma}
c^\dagger_{{\bf k}+{\bf q},\sigma}c_{{\bf k}\sigma}$. However,
matrix elements play a crucial role in the understanding of the inelastic
scattering.
Using Wannier functions, we can rewrite
the matrix element in terms of scattering from an atom at site ${\bf R}_0$
to a site ${\bf R}$,
\begin{eqnarray}
\langle \psi_{{\bf k}+{\bf q},n'\sigma} |e^{i{\bf q}\cdot{\bf r}}
|\psi_{{\bf k}n\sigma}\rangle=\sum_{\bf R} e^{i{\bf q}\cdot {\bf R}}
\langle \phi_{{\bf R}\alpha'\sigma} 
|e^{i{\bf q}\cdot {\bf r}} |\phi_{{\bf R}_0\alpha\sigma}\rangle ,
\end{eqnarray}
 where 
$\phi_{{\bf R}\alpha\sigma}$
is a localized Wannier orbital 
of type $\alpha$ at site ${\bf R}$.
The operator $\rho_{\bf q}$ can create charge-transfer transitions
(with ${\bf R}\ne {\bf R}_0$), 
plasmon excitations \cite{Soininen1,Cai,Gurtubay}, 
and dipolar and higher-order transitions from core to valence
states (where ${\bf R}= {\bf R}_0$) 
\cite{Galambosi,Hamalainen,Soininen,Sternemann}. 
Recently, it was  demonstrated  \cite{Larson} that it is also
possible to measure dipolar forbidden $dd$ transitions.
By tuning the transferred momentum \cite{Larson,Haverkort},
one is able to maximize
the intensity for  local transitions (${\bf R}_i={\bf R}_j)$
within the $3d$ shell, for which the matrix element is
\begin{eqnarray}
\langle \phi_{{\bf k}+{\bf q},n'} |e^{i{\bf q}\cdot{\bf r}}
|\phi_{{\bf k}n}\rangle
&\cong&
\langle \phi_{3d,\alpha'} |e^{i{\bf q}\cdot {\bf r}} |\phi_{3d,\alpha}\rangle, 
\end{eqnarray}
where $\alpha$ denotes the different $3d$ orbitals. This approximation assumes reasonably well 'localized' Warnier orbitals. 
These sharp dipole-forbidden transitions were first observed by
 Larson {\it et al.} \cite{Larson}  for large $q$ in NiO and CoO. 
The use of large wave vectors 
would allow a resolution of 30 meV or less, making
nonresonant IXS a promising tool to study local crystal field and
orbiton excitations. It could be used, for example, to study  the
Jahn-Teller distortions in manganites and the nature of the small
crystal field distortions in early transition-metal oxides.
Although the  angular dependence in NiO and CoO has been analyzed with
 density-functional theory \cite{Larson} and  small-cluster 
calculations \cite{Haverkort}, the detailed nature of the angular dependence
and how to use it is not well understood. 
In this paper, we express $\rho_{\bf q}$
as an effective operator, derive the selection rules governing the local
$dd$ transitions, and give explicit angular dependencies. We provide
an explanation for the remarkable intensity variations in certain directions.
The derivation for the angular distribution of the intensities is partially based on symmetry rules and should therefor be very generally valid.
We show how IXS  can be used to extract detailed ground-state 
information  by treating  the IXS from orbitons (excitonic orbital excitations not involving the Hubbard U) in manganites.

\section{ Effective operators and selection rules}
For local $dd$ transitions, it is convenient to express the 
$3d$ Wannier functions 
$\phi_{{\bf R}_0 m}= R_{3d}(r)Z^{(2)}_m ({\hat {\bf r}})$,
 with $r=|{\bf r}|$ and  ${\hat {\bf r}}={\bf r}/r$, 
in terms of a radial function  $R_{3d}(r)$ and a
real angular part 
\begin{eqnarray}
 Z^{(l)}_m({\hat {\bf r}}) = N_m \sqrt{\frac{4\pi}{2l+1}}
[Y^{(l)}_{-\overline{m}}({\hat {\bf r}})\mathopen{+}
s_m Y^{(l)}_{\overline{m}}({\hat {\bf r}})] ,
\end{eqnarray} 
where $\overline{m}=|m|$,
$N_m=1,\frac{1}{\sqrt{2}}i^{(1-{\rm sgn}m)/2}$ 
 and $s_m=0,(-1)^m  {\rm sgn} (m)$ for  $m=0$ and $\overline{m} >0$,
respectively;  $Y^{(l)}_m({\hat {\bf r}})$ is
a  spherical harmonic. The functions $Z^{(l)}_m({\hat {\bf r}})$
are known as tesseral harmonics and convenient when dealing with 
transition-metal compouns since
the values  $m=-2,-1,0,1,2$  correspond to the 
$3d$ orbitals $d_{xy}, d_{yz}, d_{3z^2-r^2}, d_{xz}, d_{x^2-y^2}$, 
respectively, see 
Table \ref{angular}. 
  We can expand the exponent in Eqn. (\ref{A2}) in terms 
of Bessel functions and tesseral  harmonics,
\begin{eqnarray}
e^{i{\bf q}\cdot {\bf r}} 
=\sum_{LM} (2L\mathopen{+}1) i^L j_L (qr) 
Z^{(L)}_{M} ({\hat {\bf q}})
Z^{(L)}_{M}({\hat {\bf r}})
\end{eqnarray}
where $M=-L,-L+1,\cdots,L$ and $j_L$ is a Bessel function of order
$L$. Spherical-tensor algebra gives an effective transition
operator
\begin{eqnarray}
\rho_{\bf q}=\sum_L A_{L}(q) \sum_{M}
Z^{(L)}_{M}({\hat {\bf q}}) 
w^L_{ M},
\label{transition}
\end{eqnarray}
consisting of a one-particle transition operator $w^L_M$ probed
by ${\bf q}$ through 
a reduced matrix element  $A_L(q)$ and an  angular dependence 
$Z^{(L)}_M({\hat {\bf q}})$.
Of the summation over $L$ only the  values 
 0 (monopolar), 2 (quadrupolar), and 4 (hexadecapolar) remain.
The factor 
\begin{eqnarray}
A_{L}(q)=i^L  (2L\mathopen{+}1) C_{20,L0}^{20} C_{22,L0}^{22} P_L,
\end{eqnarray}
where $C_{l_1m_1,l_2m_2}^{l_3m_3}$ are Clebsch-Gordan coefficients,
and $P_L(q)=\int dr r^2 R_{3d}(r) j_L(qr) R_{3d} (r)$ 
is the reduced matrix element of the Bessel function. 
For brevity, we implicitly assume the dependence on $q$
and ${\hat{\bf q}}$ in the remainder. 
$A_{L}=P_0, \frac{10}{7} P_2,\frac{3}{7}P_4$, 
for $L=0,2$, and 4, respectively.
In second quantization, the transition operator is
\begin{eqnarray}
w^L_{M}
&=&\sum_{m\sigma }
\mathop{\sum_{ m'=m_\pm}}_{m_-\ne m_+ } 
a_{mm'}^{LM}
d^\dagger_{m' \sigma}d_{m \sigma},
\label{transition}
\end{eqnarray}
where $d_{m\sigma}^\dagger$ 
creates an electron with spin $\sigma=\uparrow,\downarrow$
in the $3d$ orbital with index $m$.
The transition probability is

\begin{table}[t]
\caption{ The angular dependence  
$U_{mm'}
({\hat {\bf q}})=\langle m' |\rho_{\bf q}|m\rangle$
of the scattering between orbitals $m$ and $m'$. The real 
$3d$ orbitals are
$Z^{(2)}_{m}=\sqrt{3} \hat{x}\hat{y}$, $\sqrt{3} \hat{y}\hat{z}$,
$\frac{3}{2}\hat{z}^2-\frac{1}{2}$,
$\sqrt{3} \hat{x}\hat{z}$ ,
$\frac{1}{2}\sqrt{3} (\hat{x}^2-\hat{y}^2)$ for $m=-2,-1,0,1,2$;
${\hat {\bf q}}=(\hat{x},\hat{y},\hat{z})=(\sin\theta\cos\varphi,
\sin\theta\sin\varphi,\cos\theta)$, in conventional
spherical coordinates $\theta$ and $\varphi$. 
For off-diagonal  matrix elements ($m\ne m'$)  
for $t_{2g}$ ($m=\pm 1,-2$) orbitals,  $m''$ 
denotes the $t_{2g}$ orbital for which $m'' \ne m,m'$. The coordinate 
$\hat{r}_m=\hat{y},\hat{x},\hat{z}$ for 
$m=1,-1,\pm 2$.  
 }
\begin{ruledtabular}
\begin{tabular}{lccc}
$m $ & 
$m' $ &  $U_{mm'}({\hat {\bf q}})$  \\
\hline
$m'=m:$ \\
$0$ &   & $A_2(-\frac{3}{2}\hat{z}^2+\frac{1}{2}) +
\frac{3}{4}A_4 (35 \hat{z}^4-30 \hat{z}^2+3)$\\
$\ne 0$ &   &
$A_2(\frac{3}{2}\hat{r}_{m}^2-\frac{1}{2})+A_4(
5\hat{r}_m^2 -4 +\frac{35}{3} [Z^{(2)}_m]^2)$ \\
$m' \ne m:$ \\
$\pm 1, -2$ &  $\pm 1, -2$ &$[-\frac{\sqrt{3}}{2}A_2 
+\frac{5}{3}\sqrt{3}A_4 (7 \hat{r}_{m''}^2-1) ]
Z^{(2)}_{m''}$ \\
$\pm 1 $ & $0$ &
$[-\frac{1}{2}A_2 
+\frac{5}{2}A_4 (7\hat{z}^2-3)] Z^{(2)}_{\pm 1}$ \\
$\pm 1 $ & $2$
& $[\mp \frac{\sqrt{3}}{2}A_2 
+ \frac{5}{6}\sqrt{3}A_4\{ 7(\hat{x}^2-\hat{y}^2)\pm 2\} ] 
Z^{(2)}_{\pm 1}  $ \\
$\pm 2$ & $0$ &
$ [A_2+\frac{5}{2}A_4 (7\hat{z}^2-1)  ]Z^{(2)}_{\pm 2}$ \\
$2$ & $-2$  & $\frac{35}{2}A_4 \hat{x}\hat{y}  (\hat{x}^2-\hat{y}^2 )$
\end{tabular}
\nopagebreak
\label{angular}
\end{ruledtabular}
\end{table}
\noindent
\begin{eqnarray}
a_{mm_\pm}^{LM}
= \delta_{m_\pm ,{\rm sgn}(mM)|\overline{m}\pm \overline{M}|}
\frac{N^*_{m_\pm}  N_m N_M P^\pm_{mM}  }{  C_{20,l0}^{20}}
C_{2\overline{m},l,\pm \overline{M}}^{2,\overline{m}\pm \overline{M}}
\nonumber
\end{eqnarray}
with 
$P^-_{mM}=2 s_M, 2 s_m$ for  
$\overline{m}\pm \overline{M}\ge0$ and $\overline{m}\pm \overline{M}\le0$,  
respectively,
and $P^+_{mM}=2-\delta_{m,0}\delta_{M,0}$.

For the monopole term ($L=0$), the scattering 
is elastic and isotropic, since  
\begin{eqnarray}
w^0_{0}
=n_e=\sum_{m\sigma} d^\dagger_{m\sigma}d_{m\sigma},
\end{eqnarray}
and $Z^{(0)}_{0}({\hat {\bf q}})=1$.

{\it For inelastic scattering in transition-metal systems, the 
coefficient contains the simple selection rule }
\begin{eqnarray} 
m'={\rm sgn (mM)}|m \pm M |,
\end{eqnarray}
 under the conditions that $|m'|\le 2$
and ${\rm sgn (mM)}=1$ for $m'=0$.
This selection rule helps us to obtain an understanding of nonresonant 
IXS. As an example, let us consider  a Cu$^{2+}$ ion
in $D_{4h}$ symmetry for quadrupolar ($L=2$) scattering. 
The ground state is $|\underline{d}_{x^2-y^2}\rangle$ ($m=2$),
where the underline indicates holes.
When measuring 
along the $[001]$ direction, the only nonzero angular term
is $Z^{(2)}_{0}=\frac{3}{2}\hat{z}^2-\frac{1}{2}$ 
($m=0$), where 
we use ${\hat {\bf q}}=(\hat{x},\hat{y},\hat{z})=(\sin\theta\cos\varphi,
\sin\theta\sin\varphi,\cos\theta)$ in conventional
spherical coordinates $\theta$ and $\varphi$.  
There is no inelastic scattering, since the  
relevant transition operator $w^{2}_{0}$ only contributes to
the elastic intensity ($m=2\rightarrow 2$). When measuring with 
${\bf q}$ in the $xy$ plane,
$Z^{(L)}_M$ is zero  for odd $M$. For even $M$, 
$w^2_0$ gives elastic scattering; 
 $w^2_{-2}$ 
does not contribute since transitions to 
$m'=-0,-4$ are forbidden
since ${\rm sgn}(mM)=1$ is not satisfied for $m'=0$ and $|m'|\le 2$
for $3d$ electrons.
The only inelastic scattering is due to $w^2_{ 2}$ 
giving $\underline{d}_{x^2-y^2}$ $\rightarrow \underline{d}_{3z^2-r^2}$ 
(or $m=2\rightarrow 0$)
with a $Z^{(2)}_2=\frac{\sqrt{3}}{2}(\hat{x}^2-\hat{y}^2)$ angular dependence.
In addition, we can easily see that 
transitions  $\underline{d}_{x^2-y^2}\rightarrow \underline{d}_{xy}$ 
($m=2\rightarrow -2$) are possible for operators 
with symmetry $-0$ and $-4$ using the inverse relationship  
$M={\rm sgn}(m m')|m\pm m'|$. 
However, $-0$ does not satisfy the condition that ${\rm sgn}(mm')=1$ 
for $M=0$ and is
therefore not allowed. The transition  
$\underline{d}_{x^2-y^2}\rightarrow \underline{d}_{xy}$  is therefore 
 hexadecapolar ($L=4$, $M=-4$).

Although all the off-diagonal terms for the quadrupolar scattering are determined
by a single $M$, often two different 
 $M$ values interfere.  It is therefore convenient 
to define a total angular dependence between two $3d$ orbitals by
$U_{mm'}({\hat {\bf q}})
=\langle m' |\rho_{\bf q}|m\rangle$, which are 
given in Table \ref{angular}. We now demonstrate how the selection rules
and symmetry can help in extracting valuable ground-state information.

\begin{figure}[b]
\begin{center}
\includegraphics[angle=0,width=0.80  \linewidth]{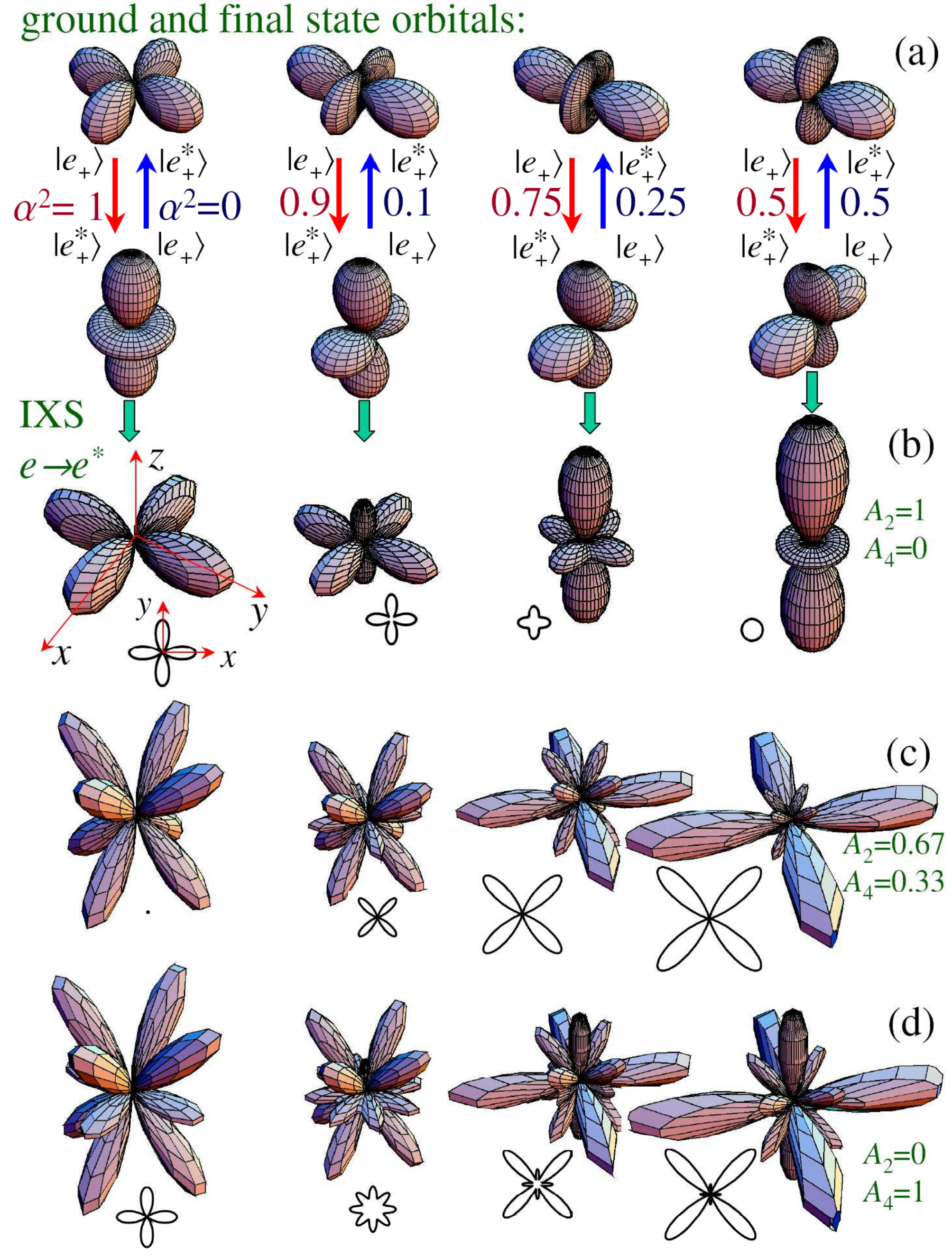}
\caption{\label{orbitonee} (color online) Inelastic scattering 
from $e\rightarrow e^*$ in the manganites. (a) The top part shows 
the lowest $e$ orbitals for 
$|e_+\rangle= \alpha |2\rangle + \beta|0\rangle$ for 
$\alpha^2=1, 0.9, 0.75$, and $0.5$. Excitations are made into 
the orbital $|e^*_+\rangle=\alpha  |0\rangle -  \beta |2\rangle$.
The angular dependence  for $\alpha^2=0, 0.1, 0.25, 0.5$ is the same as
those for $1-\alpha^2$. The inversion of the  relative energy positions of the 
orbitals does not affect the nonresonant IXS, which is 
proportional to $|\langle e^*_\pm |\rho_{\bf q}|e_\pm\rangle |^2$.
 (b) The angular distribution of the nonresonant IXS intensity in the 
quadrupolar region ($A_2=1$ and $A_4=1-A_2=0$). The intensity in a certain 
direction is proportional to the distance to the origin. The insets 
show the intensity in the $xy$ plane. The intensities give the sum 
over both orientations of the $e_\pm$ orbitals; 
(c) idem for $A_2=0.67$; (d) idem for $A_2=0$. }
\end{center}
\end{figure}

\begin{figure}[b]
\begin{center}
\includegraphics[angle=0,width=0.80  \linewidth]{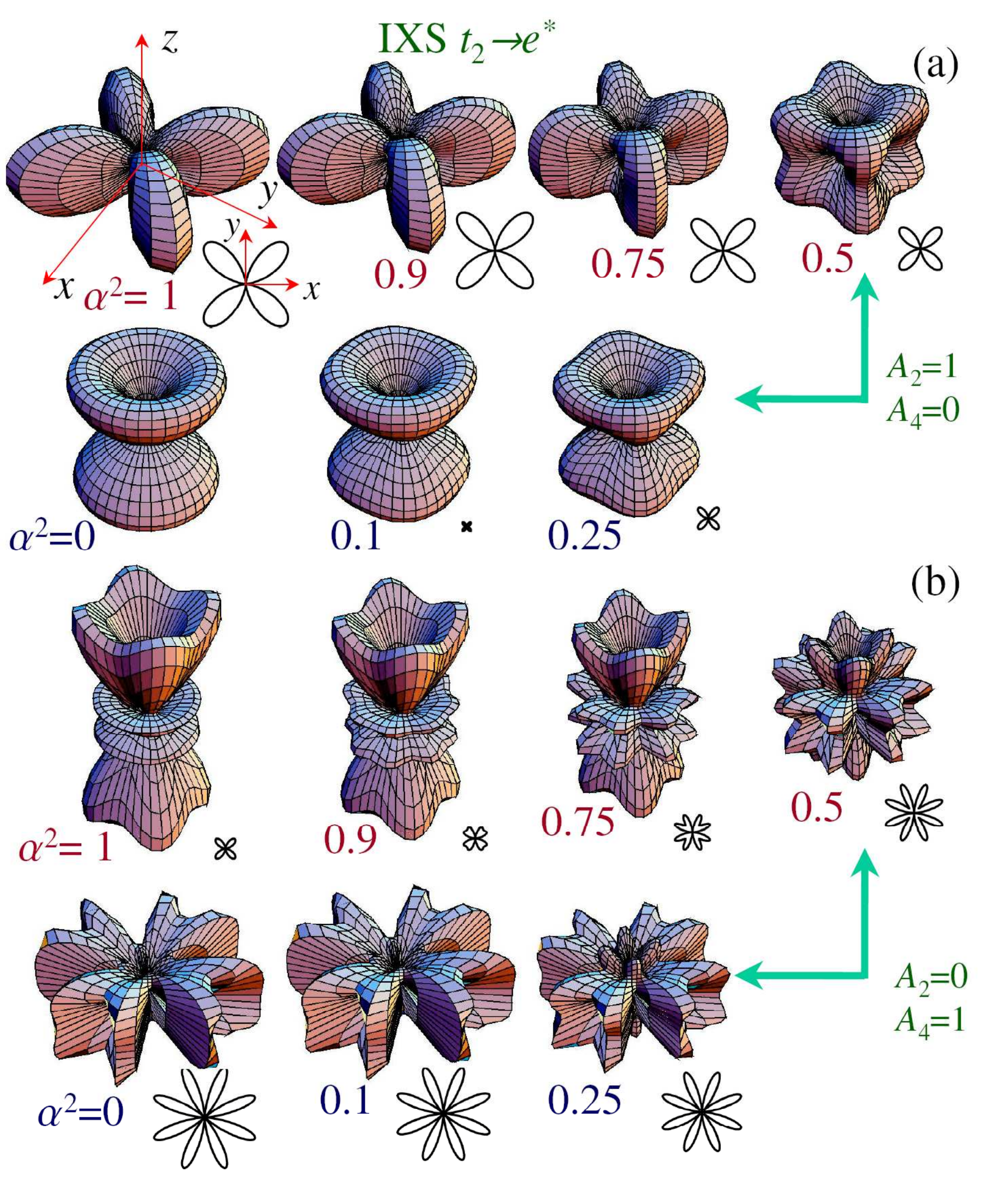}
\caption{\label{orbitonte} (color online) 
 Nonresonant inelastic scattering  for  the manganites with an electron in the 
$|e\rangle$ states as in Fig. \ref{orbitonee}(a), but now
for scattering $t_2\rightarrow e^*$  for 
$\alpha^2=1, 0.9, 0.75, 0.5, 0.25, 0.1$ and 0. Note that now $\alpha^2$
and $1-\alpha^2$ are inequivalent. (a) Nonresonant IXS in the 
quadrupolar region ($A_2=1$ and $A_4=0$);
(b) idem in the hexadecapolar region ($A_2=0$ and $A_4=1$). The insets show
the scattering intensity in the $xy$ plane. }
\end{center}
\end{figure}

\section{ IXS in high-symmetry directions}
One of the striking features of nonresonant inelastic X-ray scattering 
in NiO is the presence of a [001] nodal direction absent in CoO. 
Larson {\it et al.} \cite{Larson} ascribed
this to a ``${\bf q}$-selection rule'' associated with the nearly cubic point
group symmetry of NiO.
The absence of inelastic scattering,
$\langle f|\rho_{\bf q}|g\rangle=0$, implies 
$\rho_{\bf q}|g\rangle={\rm const}|g\rangle$.
This occurs when $H$ and 
$\rho_{\bf q}$ commute, $[H,\rho_{\bf q}]=0$. In general, this rarely 
happens, but $\rho_{\bf q}$ can commute with parts of the Hamiltonian,
which then might give rise to nodal directions for all excited states
if $\rho_{\bf q}|g\rangle={\rm const}|g\rangle$ and otherwise 
gives a nodal direction for the excited states for which 
$\langle f|\rho_{\bf q}=\langle f|{\rm const}$. Below we will first discuss
the commutation relations between $\rho_{\bf q}$ and the crystal-field
operator and then those with the Coulomb interaction.
For ${\bf q}=q {\hat {\bf z}}$, the
only nonzero angular dependence  $Z^{(L)}_{0}$ selects
 the operators $w^L_{0}$. This gives a density operator diagonal in $m$,
\begin{eqnarray}
\rho_{q {\hat {\bf z}}}
=\sum_{m\sigma} Q_m d^\dagger_{m\sigma}d_{m\sigma}
\label{rhoz}
\end{eqnarray}
with  
\begin{eqnarray}
Q_m=A_2(\frac{1}{2}m^2-1)+
A_4(\frac{35}{12} m^4-\frac{155}{12}m^2+6).
\label{Qm}
\end{eqnarray} 
In several common crystal-fields, such as $O_h$ and $D_{4h}$, $m$ is a good
quantum number and therefore $\rho_{\bf q}({\hat {\bf z}})$
commutes with the crystal-field operator. Along $[100]$, 
$\rho_{q {\hat {\bf x}}}$ is no longer diagonal but, since it is 
a unitary transformation over $90^\circ$ of $\rho_{q {\hat {\bf z}}}$, 
still only contains 
$e\rightarrow e$ and $t_2\rightarrow t_2$, but no $e\rightarrow t_2$ 
scattering. Eigenstates of the octahedral crystal field therefore scatter 
among themselves. Thus, for a system with a $O_h$ or $D_{4h}$ symmetry
a nodal direction can occur when ${\bf q}\parallel C_4 $, where
$C_4$ is a fourfold symmetry axis. Since in general $\rho_{\bf q}$ does
not commute with the Coulomb interaction, the eigenstate of the crystal
field should also be an eigenstate of the Coulomb interaction. This occurs
for high-spin $3d^n$ configurations, except $3d^2$ and $3d^7$.
For example, the ground
state for a Ni$^{2+}$ ion (in the absence of spin-orbit coupling) is 
$\underline{d}_{3z^2-r^2\uparrow} \underline{d}_{x^2-y^2 \uparrow}$.
This is an eigenstate
of  the crystal field, the Coulomb interaction,  and
$\rho_{q {\hat {\bf z}}}$. Therefore, no inelastic scattering occurs
along all six $C_4$ axes.
On the other hand,
for Co$^{2+}$, the ground state is given by
$|g\rangle=\alpha 
|\underline{d}_{t_2\uparrow}\underline{d}_{e\uparrow}^2 (^4T_1)\rangle
+\beta 
|\underline{d}_{t_2\uparrow}^2 \underline{d}_{e\uparrow} (^4T_1)\rangle$, 
where the mixing occurs due the Coulomb interaction.
This is not an eigenstate of the octahedral crystal field and the operators
$\rho_{q {\hat {\bf z}}}$, and inelastic scattering in the [001] direction
therefore occurs into the excited multiplet 
\begin{eqnarray} 
|f\rangle=\beta
|\underline{d}_{t_2\uparrow}\underline{d}_{e\uparrow}^2 (^4T_1)\rangle
-\alpha
|\underline{d}_{t_2\uparrow}^2 \underline{d}_{e\uparrow} (^4T_1)\rangle,
\end{eqnarray} 
which is about 
2.4 eV higher in energy \cite{Haverkort}.  Also, deviations from a
 $t_2e^2$ ground state due to band effects and a
lowering of the crystal field \cite{Larson}  give rise to inelastic 
scattering. Spin-orbit coupling can also remove a nodal direction,
but its effect is small except when the spin-orbit coupling lifts 
a degeneracy. 
Finally, the conditions for nodal directions above 
are generally not satified for 
low- and intermediate-spin ground states.

\section{ Orbiton excitations in manganites}
Orbital physics plays an important role in manganites which have 
been extensively studied for their magnetoresistive behavior \cite{Salamon}.
LaMnO$_3$ is known to have an alternating $d_{3x^2-r^2}/d_{3y^2-r^2}$
orbital ordering, and half-filled systems often display the $CE$ type structure
with charge and orbital order occurring in an unconventional zigzag magnetic
structure. However, others have contested this ionic picture stating that
the ground state is more complex. For example, for 
La$_{0.5}$Sr$_{1.5}$MnO$_4$, it was claimed, based 
on X-ray magnetic linear dichroism experiments \cite{Huang}, that a significant
out-of-plane character was mixed in, giving an orbital ordering 
close to $d_{x^2-z^2}/d_{y^2-z^2}$. We demonstrate  the extreme sensitivity 
of the angular dependence of nonresonant IXS on the detailed nature of the 
ground state. The ground state for Mn$^{3+}$ is given by a 
$d^3_{t_2\uparrow}d_{e\uparrow}$ configuration. Degenerate $e$ orbitals
$d_{3z^2-r^2}$ ($m=0$) and $d_{x^2-y^2}$ ($m=2$)  
are sensitive to distortions
leading to lowest states given by
$|e_\pm\rangle= \pm \alpha |2\rangle + \beta|0\rangle$,
with $\alpha^2+\beta^2=1$. The  phases account for the different 
orientations of the orbitals due to orbital ordering 
often found in manganites.
 With nonresonant IXS, one can make excitations into the 
empty state $|e^*_\pm\rangle=\alpha  |0\rangle \mp  \beta |2\rangle$
by exciting an electron from the $e$  ($d_{t_2\uparrow}^3d_{e_\pm \uparrow}
\rightarrow d_{t_2 \uparrow}^3 d_{e^*_\pm \uparrow}$)
or $t_2$ states ($d_{t_2 \uparrow}^3d_{e_\pm \uparrow}
\rightarrow d_{t_2\uparrow}^2d_{e_\pm \uparrow} 
d_{e_\pm^* \uparrow}$). 
For the former, the inelastic intensity summed over both 
 $e_\pm$ orientations is
\begin{eqnarray}
\sum_{p=\pm} |\langle e_p^*|\rho_{\bf q}|e_p\rangle |^2 &=&
2\alpha^2 \beta^2 (U_{00}\mathopen{-}U_{22})^2 \nonumber 
+2(\alpha^2\mathopen{-}\beta^2)^2 U_{20}^2, \nonumber
\end{eqnarray}
see also Table \ref{angular}.
Figure \ref{orbitonee} shows the angular dependencies for different
ground-state orbitals given by $\alpha^2$, see Fig. \ref{orbitonee}(a). 
Let us first look at the
quadrupolar region ($A_2=1$, $A_4=0$), see Fig.
\ref{orbitonee}(b). The angular dependence
can be rewritten as $8\alpha^2 \beta^2( \frac{3}{2}\hat{z}^2-\frac{1}{2})^2+
2(\alpha^2-\beta^2)^2 \frac{3}{4} (\hat{x}^2-\hat{y}^2)^2$ and 
can be straightforwardly used to extract the value of $\alpha$.
A typical experiment is comparable to those of Larson {\it et al.}
\cite{Larson}. One fixes the magnitude of the transferred momentum and changes
its angle. For example, the intensity when 
rotating the angle from the $[001]$ to the
$[100]$ direction depends strikingly on $\alpha$.
For $\alpha=1$, which corresponds to an electron in the $x^2-y^2$ orbital,
the IXS intensity is predominantly along the $x$ and $y$ directions.
For $\alpha=\frac{1}{\sqrt{2}}$, the angular intensity is predominantly 
along the $z$ direction. 
However, when the hexadecapolar contribution increases, such an 
interpretation is less straightforward, see Fig. \ref{orbitonee}(c) and 
(d). 
However, simplifications
occur when looking in specific directions or planes. First, note
that the $[001]$ direction is only nodal when $|e_+\rangle=|2\rangle$
or $|0\rangle$, which are good eigenfunctions of $\rho_{q {\hat {\bf z}}}$.
However, more quantitative information can be obtained by
comparing the intensities along $[001]$ and $[100]$ or $[010]$. 
The $e\rightarrow e$ scattering along $[001]$ depends on the parameters 
$Q_0=-A_2+6A_4$ and $Q_2=A_2+A_4$, see Eqns. (\ref{rhoz})
and (\ref{Qm}). 
However, for inelastic scattering, 
we can remove one of those parameters by 
rewriting the  scattering operator as
 $\rho_{q {\hat {\bf z}}}=Q_0 n_e+\sum_{m\ne 0,\sigma} (Q_m-Q_0) 
d_{m\sigma}^\dagger d_{m\sigma}$. The number operator
$n_e$  only gives elastic scattering, so effectively
the $e\rightarrow e$ scattering only depends on $Q_{20}=Q_2-Q_0=2A_2-5A_4$. 
Along $[100]$, 
$\rho_{q {\hat {\bf x}}}$ only contains 
$e\rightarrow e$ and $t_2\rightarrow t_2$, but no $e\rightarrow t_2$ 
scattering.
 The $e\rightarrow e$ scattering again only depends on 
$Q_{20}$.
A straightforward calculation gives for the ratio of the nonresonant
scattering intensities in the $[001]$ and $[100]$ direction
\begin{eqnarray} 
I_{[001]}/I_{[100]}=16 \alpha^2\beta^2/(3-8\alpha^2\beta^2).
\end{eqnarray} 
Note that, since $\beta^2=1-\alpha^2$, this ratio is independent of the
reduced matrix elements and can be used to extract directly the nature
of the Jahn-Teller distorted state. From the factor  
$Q_{20}=2A_2-5A_4$, one also sees
 a destructive interference independent of $\alpha$ 
between quadrupolar and hexadecapolar
terms in the $x$, $y$, and $z$ directions, which 
 is clearly visible in Fig. \ref{orbitonee}(c), where the intensity 
in those direction is almost zero.
A clear feature that displays the change in ground state as a 
function of $\alpha^2$ is the direction of the dominant lobes 
in the $xy$ plane, see insets in Fig. \ref{orbitonee}(b)-(d).
Again, this can be used to obtain information on ground-state properties.
The intensity in the $x$ or $y$ direction 
$\frac{1}{8}Q_{20}^2 (3-8 \alpha^2\beta^2)$ decreases with increased
mixing of the $x^2-y^2$ and $3z^2-r^2$ orbitals,
whereas along the $[110]$ directions, 
the intensity $2\alpha^2\beta^2 (A_2+\frac{25}{4}A_4 )^2 $
increases. 

Although the measurement of the $e\rightarrow e^*$ orbiton excitation
has the great advantage that the energy loss can be directly related to 
the energy for an orbiton excitation.
Since this energy is of the order of 0.3-1.5 eV, 
this feature might in practice 
be difficult to distinguish from the elastic line. However, valuable 
information can still be obtained from the study of the $t_2\rightarrow e^*$
excitation, where an electron from the $t_{2g}^3$ spin is excited into
the empty $e^*$ orbital.
This excitation is at higher energy loss due to the 
crystal-field splitting between $e$ and $t_2$, which is 
of the order of 2-2.5 eV. The angular dependencies
are shown in Figure \ref{orbitonte}. First, note that the spectra for
$\alpha^2$ and $1-\alpha^2$ are no longer equivalent. Second, 
the $x$, $y$, and $z$ directions are all nodal since no $t_2\rightarrow e$
scattering occurs. Simplifications also occur in the $xy$ plane. Scattering
from $|0\rangle\rightarrow|\mathopen{\pm} 1\rangle$,  $|-2\rangle$ requires
$M=\pm 1,-2$ and 
$|2\rangle\rightarrow |\mathopen{\pm}1\rangle$,  $|-2\rangle$ requires
$M=\pm 1(\pm 3)$ and $(-4)$, respectively ($M$-values in parentheses give
pure hexadecapolar terms). All the odd $M$ are zero in the $xy$ plane
and only $|0\rangle\rightarrow |\mathopen{-}2\rangle$ ($M=-2$)
and $|2\rangle\rightarrow |\mathopen{-}2\rangle$ ($M=-4$) 
remain, giving an angular distribution in the $xy$ plane
$2\alpha^2 (U_{0,-2})^2+2\beta^2(U_{2,-2})^2$. 
The relative strengths of
$U_{0,-2}(\hat{x},\hat{y},0)=\frac{\sqrt{3}}{2}Q_{20}\hat{x}\hat{y}$
and $U_{2,-2}(\hat{x},\hat{y},0)=\frac{35}{2}A_4 \hat{x}\hat{y} (
\hat{x}^2-\hat{y}^2)$ 
reflect the amount of $3z^2-r^2$ and $x^2-y^2$ character in the ground state,
respectively.  Note that in the quadrupolar region, we have a 
simple scaling of the $\hat{x}\hat{y}$ dependence, 
see Fig. \ref{orbitonte}(a). The advantage of
using hexadecapolar excitations, 
see Fig. \ref{orbitonte}(b), is the clear change in angular dependence from
a four-lobed to an eight-lobed shape as a function of $\alpha$.

\section{ Summary} The nonresonant inelastic X-ray scattering for local 
$dd$ transitions has been analyzed. The strong sensitivity of the angular
dependence on the detailed nature of the ground state in combination with
the experimental degrees of freedom (scattering angle, incoming energy) and
the possible high resolution make nonresonant IXS a powerful tool to 
study crystal-field and orbital excitations. Future theoretical work should
include an analysis for the rare-earths.

\section{ Acknowledgments}
We acknowledge George Sawatzky and Hao Tjeng for useful discussion.
This work was  supported by 
 the U.S. Department of Energy (DOE), DE-FG02-03ER46097, and NIU's Institute
for Nanoscience, Engineering, and Technology under a grant from the U.S.
Department of Education. Work at
Argonne National Laboratory was supported by the U.S. DOE, 
Office of Science, Office of Basic Energy Sciences, under contract 
DE-AC02-06CH11357. The work in Cologne was supported by the Deutsche 
Forschungsgemeinschaft through SFB 608.


\begin{thebibliography}{99}
\bibitem[*]{ESRF} Part of this work was done during MvV's stay at
 the European Synchrotron Radiation Facility.
\end{thebibliography}
\end{document}